\documentclass[trackchanges]{aastex7}

\usepackage{amsmath}
\usepackage[normalem]{ulem}
\usepackage[capitalize]{cleveref}
\usepackage[caption=false]{subfig}
\usepackage{comment}
\usepackage{flushend}
\usepackage[T1]{fontenc} 
\usepackage{mathtools}
\usepackage{xcolor}
\usepackage{graphicx}
\usepackage{dcolumn}
\usepackage{bm}
\usepackage{multirow}
\usepackage{slashed}

\usepackage{epstopdf}

\newcommand{\Beq}{\begin{equation}\begin{aligned}}
\newcommand{\Eeq}{\end{aligned}\end{equation}}

\begin{document}

\preprint{KEK-QUP-2024-0027, KEK-TH-2675, KEK-Cosmo-0368}

\title{Diffuse Neutrino Signals from Dark Stars Seeding Super-Massive Black Holes}

\author[orcid=0000-0002-8471-6879]{Thomas Schwemberger}
\affiliation{Department of Physics and Institute for Fundamental Science, University of Oregon, Eugene, OR 97403, USA}
\email{tschwem2@uoregon.edu}

\author[orcid=0000-0003-2647-3431]{Volodymyr Takhistov}
\email{vtakhist@post.kek.jp}
\affiliation{International Center for Quantum-field Measurement Systems for Studies of the Universe and Particles (QUP, WPI),
High Energy Accelerator Research Organization (KEK), Oho 1-1, Tsukuba, Ibaraki 305-0801, Japan}
\affiliation{Theory Center, Institute of Particle and Nuclear Studies (IPNS),  High Energy Accelerator Research Organization (KEK), Tsukuba 305-0801, Japan}
\affiliation{Graduate University for Advanced Studies (SOKENDAI), \\
1-1 Oho, Tsukuba, Ibaraki 305-0801, Japan}
\affiliation{Kavli Institute for the Physics and Mathematics of the Universe (WPI),   The University of Tokyo Institutes for Advanced Study, The University of Tokyo,  Kashiwa, Chiba 277-8583, Japan}

\begin{abstract}
Dark stars (DSs) — first stars powered by dark-matter (DM) heating rather than fusion—could form in the early Universe.
They can grow to $\gtrsim 10^5 M_{\odot}$ masses and collapse into seeds of supermassive black holes (SMBHs). 
We demonstrate that diffuse neutrino flux generated by DSs can be observable in existing experiments and have energies reaching hundreds of MeV, providing novel window for probing SMBH progenitors.
We establish first constraints on DSs and DM annihilations powering them using data from Super-Kamiokande and IceCube neutrino experiments, and consistent with James Webb Space Telescope observations.
Upcoming experiments such as Hyper-Kamiokande, DUNE, JUNO will be able to explore DS properties with enhanced sensitivity.
\end{abstract}

\section{Introduction}
The first generation stars are thought to have played a pivotal role in the Universe, including shaping cosmic evolution, influencing reionization and chemical enrichment. These population III (Pop III) zero-metallicity stars are believed to have formed from the gravitational collapse of pristine gas left over after the Big Bang, within dark matter (DM) mini-halos of mass
$\sim 10^5-10^7 M_{\odot}$ around redshifts $z \sim 20$~\citep{Abel:2001pr,Bromm:2001bi,Yoshida:2006bz}.
Despite their significance, Pop III stars are yet to be observed, although candidates have been suggested~\citep{vanzella:2020,welch:2022,wang:2024}.

Supermassive black holes (SMBHs) with masses ranging $\sim 10^6 - 10^9 M_{\odot}$  are ubiquitous at the centers of galaxies, powering quasars and active galactic nuclei (AGN). The origin of SMBHs at high redshifts remains a mystery, especially in light of recent observations by the James Webb Space Telescope (JWST)~\citep{Matthee:2023utn,Yue_2024,Ding:2023,stone:2024} suggesting SMBH detections even at redshifts of up to 
$z\simeq10.3$~\citep{Bogdan:2023ilu}. Achieving the necessary rapid growth of such massive objects challenges Eddington-limited accretion and hierarchical mergers of stellar-mass black holes on relevant cosmological timescales~\citep{Venemans:2013npa,Banados:2017unc}. While Pop III stars offer plausible candidates for forming lighter $\sim 10^2-10^3 M_{\odot}$ SMBH seeds, alternative scenarios such as direct collapse have also been suggested~\citep{Volonteri:2010wz,Inayoshi:2019fun} and further investigation is necessary.

An intriguing distinct class of first stars, known as dark stars (DSs), powered by heating from DM constituting a small fraction of their mass instead of fusion, has been put forth and that could seed SMBHs~\citep{Spolyar:2007qv,Freese:2008wh,Spolyar:2009nt}. DM heating can be a general byproduct of annihilation of weakly interacting massive particles (WIMPs)~\citep{Spolyar:2007qv}, but can also arise in contexts such as self-interacting DM~\citep{Wu:2022wzw}. During formation, the contraction of baryonic matter induces adiabatic contraction of DM, leading to high DM densities sufficient for efficient annihilation. The energy released from DM annihilation halts the collapse at low temperatures, preventing the onset of nuclear fusion. When the DM fuel is exhausted, the DS collapses into a black hole. The formation and sustained growth of DSs is a topic of active research~\citep{Ripamonti:2010ab, Sivertsson:2010zm, Iocco:2008rb, Smith:2012ng, Stacy:2013xwa}. Massive DSs seeding SMBHs could also explain paucity of intermediate-mass black holes~\citep{tan2024}.

Although detection of DS is challenging, their electromagnetic emission contributing to photon background has been considered~\citep{Schleicher:2008gk,Maurer:2012fv,Yuan:2011yb}.
Recently, it has been suggested that some high-redshift galaxy candidates observed by JWST could be spectroscopic manifestations of DSs~\citep{Ilie:2023zfv}, although this has been debated~\citep{Iocco:2024rez}. 

In this work we investigate neutrino signatures originating from DSs and establish a new observational window into the progenitors of SMBHs. Going beyond previous limited considerations~\citep{Schleicher:2008gk,Iocco:2008xb}, we show that DSs can produce detectable neutrino signals. The DS neutrino flux can be orders of magnitude larger than expected from gamma ray bursts induced by collapses of Pop III stars~\citep{Iocco:2007td, PhysRevD.85.123003} and with neutrino energies significantly larger than typical emission associated with Pop III star lifetimes~\citep{Iocco:2004wd}. We set first constraints on DM interactions fueling DS and their existence using data from neutrino experiments. Furthermore, we show that our findings can be consistent with observations from JWST complementary to neutrinos. 

Our study bridges gaps between other directions, such as neutrino
signatures~\citep{Shi:1998nd,Shi:1998jx,Linke:2001mq,Munoz:2021sad} from intermediate supermassive stars that might arise from star cluster collapse
or primordial clouds in astrophysical SMBH formation paths.
Our work is also complementary to potential neutrino signatures that could arise from DM overdensity ``spikes'' around massive black holes~\citep{Freese:2022ouh}. This provides a new approach for probing the possible nature of earliest stars.

\section{Dark star population}
The DS population is expected to trace that of the first DM halos.  
The comoving number density distribution of DM halos based on a generalized halo mass function can be expressed as
\begin{equation}\label{eq:HFR} 
    \frac{dn_h}{dM_h} = \frac{\rho_m}{M_h}\sqrt{\frac{2 A^2 \beta}{\pi}}\Big(1+(\beta \nu^2)^{-p}\Big) e^{- \beta\nu^2/2} \dfrac{d\nu}{dM_h}~,
\end{equation}
where $\nu = \delta_c/D(z)\sigma(M_h)$ with $D(z)=\delta(a)/\delta(1)$ being the linear growth factor~\citep{growth_factor}, $a = (1+z)^{-1}$ being the scale factor, and $\delta$ as defined in~Ref.~\citep{1992ARA&A..30..499C}. Here, $\delta_c=1.68$ is the critical overdensity above which a spherically symmetric perturbation region will collapse to form virialized halo. 
$\rho_m=\rho_c \Omega_m(1+z)^3$ is the average mass density of the Universe scaled to critical density $\rho_c =9\times10^{-30}$~g cm$^{-3}$ at present by fraction $\Omega_m=0.3$~\citep{Planck:2018vyg}. In Eq.~\eqref{eq:HFR} parameters $(A, \beta, p) = (0.5,1,0)$ correspond to the Press-Schechter mass function~\citep{1974ApJ...187..425P}, which we focus on throughout. We find similar results for the Sheth-Tormen~\citep{Sheth:1999mn} mass function, corresponding to $(A, \beta, p) = (0.322, 0.707, 0.3)$. The variance of the initial density fluctuation field $\sigma(M_h)$ is 
\begin{equation}
    \sigma^2(M_h) = \frac{1}{2\pi^2}\int W^2(kR_{M_h}P_\delta(k)k^2dk
\end{equation}
where $W(x) = 3 (\sin x - x \cos x)/x^3$ is a top-hat window function smoothing density fluctuation over scale $R_{M_h} = (3 M_h/4 \pi \rho_m)$. This is derived from the matter power spectrum $P_\delta(k) = A_s (k~{\rm Mpc}^{n_s} T^2(k)$ where $A_s$ is normalized by $\sigma_8$~\citep{Pierpaoli:2000ip} and $T(k)$ is transfer function fitted to cold DM model~\citep{1986ApJ...304...15B} with a spectral index $n_s = 0.965$~\citep{Planck:2018vyg}.

We model the DS formation rate to be proportional to the halo formation rate $d^2n_h/dM_hdt$ up to a scaling $f_\textrm{SMDS}$ that quantifies the fraction of DM halos hosting DSs. We do not consider the possible dependence of $f_\textrm{SMDS}$ on halo mass and redshift and assume minimal delay between halo formation and the DS formation of initial mass $M_{{\rm DS},i} \simeq 5~M_\odot$, which doesn't significantly affect our results unless these timescales are comparable to the DS lifetime. The DS growth proceeds by accretion, 
and we consider that halo mass does not appreciably changes during this period. 
We have confirmed that our DS luminosity as a function of mass matches
numerically computed 1D stellar evolution of DS results of Ref.~\citep{Rindler-Daller:2014uja} within a factor of few, assuming a constant accretion rate of
$\dot{M}_{\rm DS} = 10^{-9} M_h$~yr$^{-1}$ with DS mass following $M_{\rm DS}(t) = M_{{\rm DS},i} + \dot{M}_{\rm DS}t$. 

The comoving number density of DSs surviving to a redshift $z$ at a given cosmological time $t$ in halos of mass $M_h$ and age $\tau$ is given by
\begin{equation}\label{eq:mass_func} 
    \frac{dn_{\rm DS}}{dM_h}  = f_\textrm{SMDS}  \int_{z(t)}^{\infty} dz'\left|\frac{dt}{dz'}\right| \frac{d^2n_h}{dtdM_h}\left(M_h, z'(t-\tau)\right)~,  
\end{equation}
where we consider $M_{\rm DS}^{\rm lim} = 10^{-2} M_h$ as the upper limit of DS mass. Here $dz/dt = (1+z) H(z) $ with the Hubble parameter for $\Lambda$CDM $H(z) \simeq H_0 \sqrt{\Omega_m(1+z)^3 + \Omega_\Lambda}~ $, where $H_0 = 67.4$~km/s/Mpc is the Hubble constant at present, $\Omega_\Lambda=0.7$ accounts for dark energy density and the radiation density of the Universe is neglected.

The total DS number $n_{\rm DS}$ can then be obtained by integrating Eq.~\eqref{eq:mass_func} over the halos with DSs. For the lower limit we take $M_{h, {\rm min}}\sim10^6~M_\odot$ as the minimum mass where molecular hydrogen cooling causes protostellar clouds to collapse~\citep{ilie_observing_2012}. If lighter halos $M_h \sim 10^5~M_\odot$ allow for DSs, this could significantly enhance their abundance (see Appendix). However, we estimate that the lighter and  less-luminous DSs would only contribute a factor $\sim2$ enhancement to the total resulting neutrino flux. For the upper halo mass limit we take $M_{h, {\rm max}}\sim10^9~M_\odot$, but our results are not very sensitive this limit as such massive halos are rare at $z\gtrsim 15$.

The exact lifetimes and upper mass limits of DSs require further exploration. A large upper limit on DS mass being $\sim5\%$ of the halo mass was considered in~Ref.~\citep{ilie_observing_2012}. 
Recent high redshift SMBH observations appear to favor SMBH formation from $\sim 10^5~M_\odot$  ``heavy seeds'' at $z\gtrsim15$~\citep{Bogdan:2023ilu}. Considering halos of mass $M_h>10^6~M_\odot$ as the supermassive DS hosts, the average halo mass is around $\sim 10^7~M_\odot$. Typical DS masses of $\sim 10^5~M_\odot$ can be achieved considering DSs grow to $\sim 1\%$ of host halo mass. Our results can be readily rescaled to account for different DS masses by appropriately considering the DM cross-section in Eq.~\eqref{eq:Lum_approx}. 
Smaller DSs are more challenging to detect with telescopes, with weakened constraints and allowing for higher abundance at lower redshifts. 

Considering DS formation in mini-halos at redshifts $z \gtrsim 25$ and that they undergo collapse by $z \simeq 15$, as discussed below, the corresponding DS lifetime is $\mathcal{O}(10^8)$ yrs. Such DSs powered to luminosities of $\sim 10^{10}~L_\odot$ will require fueling of $\sim 10^6~M_\odot$ of DM, roughly percent-level of typical halo masses. 
DSs consuming heavier DM have smaller luminosities and can be expected to deplete host halos less significantly~\citep{Freese_2010}.

\section{Dark star emission}
As baryonic matter undergoes protostellar collapse, it deepens the gravitational potential well at the center of a DM mini-halo of mass $M_h$. This alters the orbits of dissipation-less DM particles in accordance with the adiabatic invariant~\citep{1986ApJ}, $r M(r) = {\rm const.}$, where $M(r)$ is the mass enclosed within radius $r$. As 
$M(r)$ increases due to baryonic inflow, $r$ must decrease, leading to the adiabatic contraction of DM. This process concentrates DM at the center of the halo, where it can reach sufficiently high densities for efficient annihilation~\citep{Spolyar:2007qv}. The energy released by DM annihilation heats the surrounding baryonic gas, stalling further collapse. The additional energy source in stellar evolution results in formation of a new DS category of early stars with much lower densities and temperatures.

We compute DS luminosity evolution considering an analytic polytropic model as described in detail in Appendix~\ref{app:polytrope}, see also~\citep{Freese_2008, Freese_2008_capture, Freese:2008hb}. The resulting luminosity for DS of mass $M_{\rm DS}$ can be well fit by
\begin{equation}\label{eq:Lum_approx}  
    L_{\rm DS} 
    \simeq ~ 2.1 \times 10^4  \left(\frac{M_{\rm DS}(t)}{M_\odot}\right)^{0.85}  \left(\frac{\langle \sigma v \rangle}{10^{-26}~\textrm{cm}^2/\textrm{s}}\right)^{0.45}\notag \times \left(\frac{m_{\chi}  (1-f_\nu)}{100~\textrm{GeV}}\right)^{-0.46} L_\odot~,
    \end{equation}
where $\langle \sigma v \rangle \simeq 10^{-26} {\rm cm}^3$s$^{-1}$ is the reference thermally averaged DM annihilation
cross-section as motivated by thermal relic abundance~\citep{relic_cross_sec}, $f_\nu$ is the fraction of the DM mass $m_{\chi}$ that is converted to neutrino energy. We find approximate agreement with results of~\citep{Freese_2010}.

\begin{figure}
    \centering
    \includegraphics[width=0.45\linewidth]{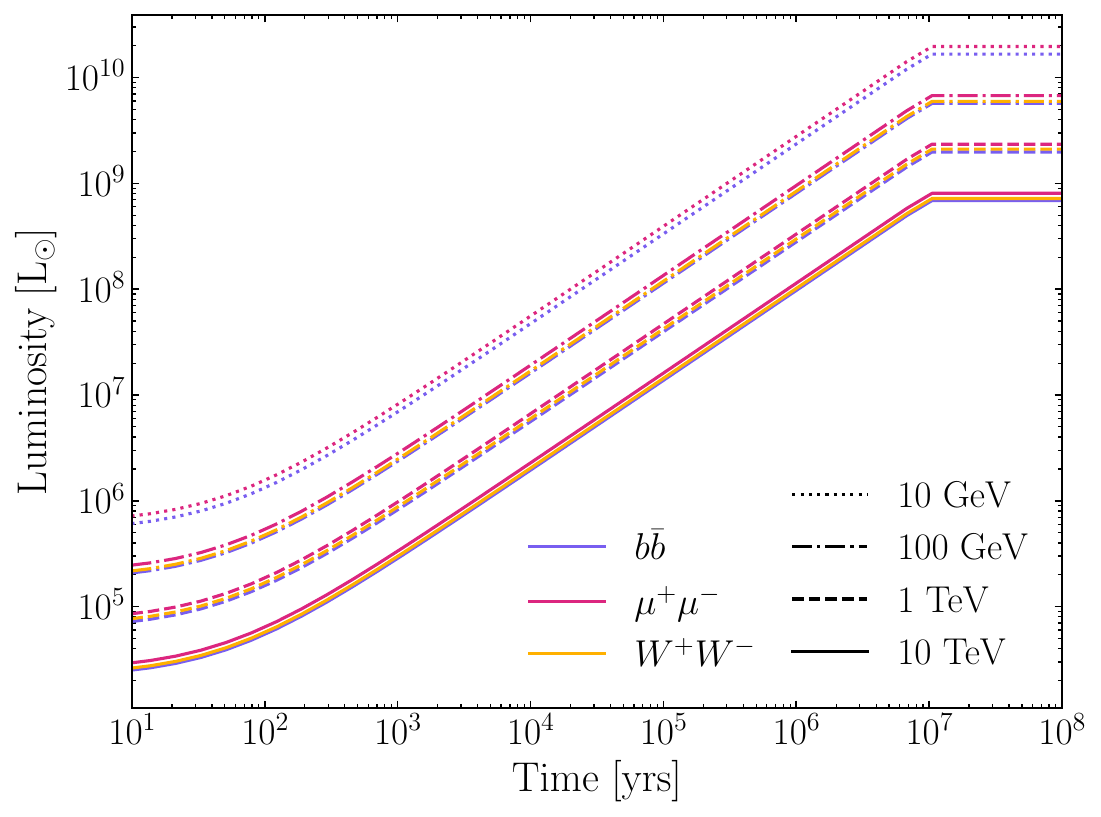}
    \caption{Evolution of DS luminosity for different DM masses and annihilation channels. We assume a baryonic accretion rate of $0.1M_\odot$~yr$^{-1}$ in a $10^8M_\odot$ halo until DS reaches a mass of $\sim 10^6M_\odot$. We consider that baryonic accretion stalls when DS reaches approximately $\sim 1\%$ of the host halo mass, resulting in a plateau corresponding to DS mass and luminosity being approximately constant until DM annihilation becomes inefficient and DS collapses.}
    \label{fig:DS_evo}
\end{figure}

The much lower temperatures of DSs, with $\lesssim \textrm{few}\times10^4$~K~\citep{Rindler-Daller:2014uja} at their surface, compared to Pop III stars result in lack of strong stellar winds. This facilitates persistent accretion of surrounding gas~\citep{Freese:2015mta}. Depending on the mass of the host halo, DM concentration as well as baryonic accretion rates DSs can reach masses of $\sim 10^{6-7}M_\odot$ and luminosities of $\sim 10^{10}L_\odot$~\citep{Freese_2016}. In Fig.~\ref{fig:DS_evo} we display the evolution of luminosities of DSs that grow to $\sim 10^6~M_\odot$ for different DM masses and annihilation channels, assuming $\dot{M}_{\rm DS} = 0.1 M_{\odot} $~yr$^{-1}$ accretion rate in a $M_h = 10^8 M_{\odot}$ halo and that DM fueling persists for extended period resulting in approximately constant DS mass and luminosity once the star reaches $M_{\rm DS} \sim 10^{-2} M_h$. This constant luminosity period dominates the total emission. Our model does not rely on assuming DM capture through baryon interactions, with DS fueling occurring through centrophilic orbits that pass through the star as discussed in Appendix~\ref{app:polytrope}.

When DSs reach their maximal mass we consider efficient accretion to stall, leading to a phase of sustained stability. This persists until the surrounding DM in the host halo is sufficiently depleted, causing DM annihilation—the primary energy source of DSs—to become inefficient. At this stage, DSs collapse. High-mass DSs could collapse directly into black holes, though a brief intermediate phase of nuclear fusion is possible~\citep{Freese_2016}. 

\section{Supermassive black hole progenitors and JWST}
Supermassive DSs constitute intriguing possible early massive $\sim 10^{4-6} M_{\odot}$ seeds for SMBHs. We consider DS number densities consistent with distribution of SMBHs at galactic centers, one per $10^{10} M_{\odot}$ baryon mass object as can be found in some dwarf galaxies,
analogous to the method of Ref.~\citep{Munoz:2021sad}.
For this, we set $f_\textrm{SMDS}$ in Eq.~\eqref{eq:mass_func} such that number density of SMBH seeds is $n_\textrm{SMBH}(z)/(1+z)^3 \sim \rho_B/10^{10} M_\odot $, where $n_\textrm{SMBH}(z)$ is given by integrating Eq.~\eqref{eq:mass_func} for contributing halos $M_h=10^6$ to $10^9~M_\odot$ and $\rho_B= \rho_c  \Omega_b = 4\times10^{-31}$g/cm$^3$ is the baryon density of the Universe with $\Omega_b=0.0476$~\citep{Planck:2018vyg}. We then evaluate $n_\textrm{SMBH}(z)/(1+z)^3$ just before DS collapse, taking $z=15$ for reference, and compare with $\rho_B/10^{10} M_\odot$ at present.
We find $f_\textrm{SMDS}\simeq 6 \times 10^{-3}$. This estimate is conservative as it does not account for such effects as mergers. 
To account for additional uncertainties, we consider $f_\textrm{SMDS}$ in the range from 1\% to 0.1\%. Additional precision in $f_\textrm{SMDS}$ evaluation can be obtained from 
detailed simulations of structure growth beyond our scope, including feedback effects~\citep{tan2024}.

Supermassive DSs could be directly detected by sensitive telescopes~\citep{ilie_observing_2012, Zackrisson2010,Ilie:2023zfv}, providing an independent and complimentary probe. In particular, JWST's infrared sensitivity, ability to probe the high-redshift Universe and spectroscopic tools uniquely position it to search for DSs. Recently, three supermassive DS candidates at redshifts $z \sim 11-14$ have been claimed from JWST Advanced Deep Extragalactic Survey (JADES) data~\citep{Ilie:2023zfv}. 
Ref.~\citep{ilie_observing_2012} used Hubble Space Telescope (HST) data to constrain DSs up surviving to redshift $z = 10$, and showed that JWST can efficiently detect supermassive DSs up to $z=15$ for $M_{\rm DS} = 10^{6}~M_\odot$.

\begin{figure*}[t]
\subfloat{\includegraphics[width=.48\linewidth]{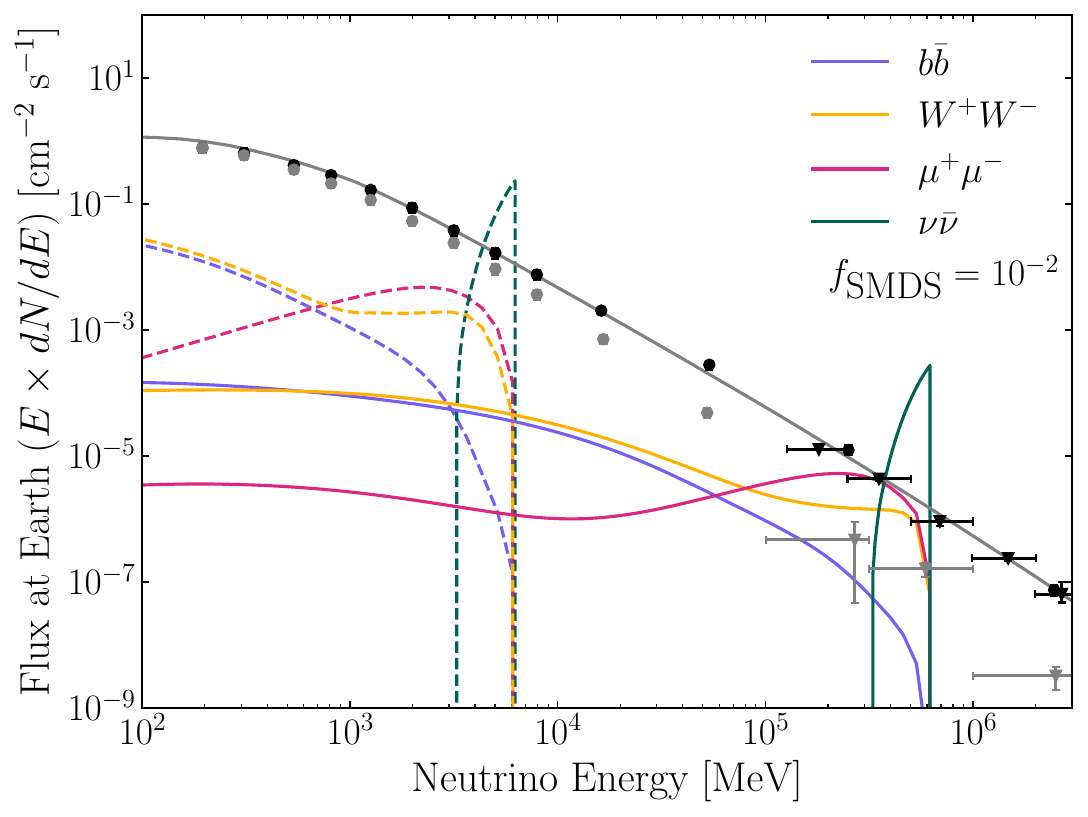}}\hfill
\subfloat{\includegraphics[width=.49\linewidth]{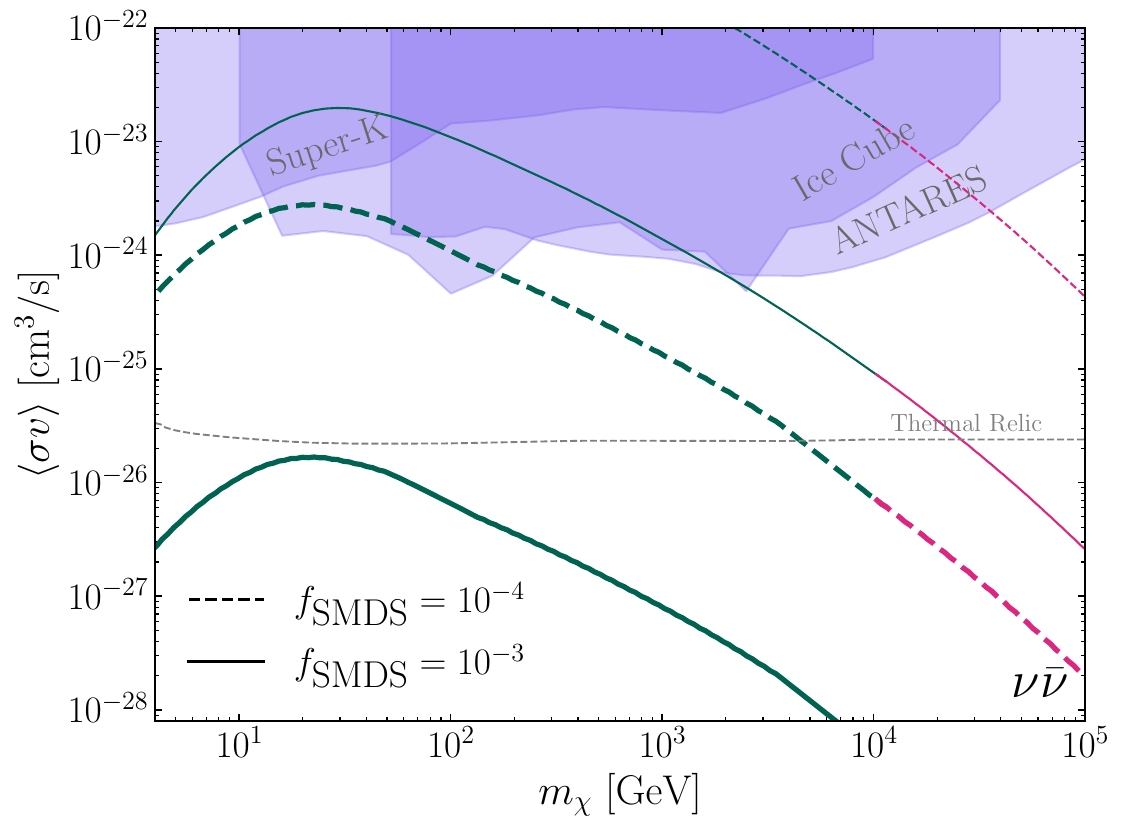}}
\caption{[Left] The neutrino flux from $10^6~M_\odot$ supermassive DSs that can seed SMBHs for different DM annihilation channels considering DM mass $m_{\chi} = 100$~GeV (dashed line) and 10 TeV (solid line), for reference cross-section $\langle \sigma v\rangle = 3\times10^{-26}$. Measurements of atmospheric neutrino flux by Super-Kamiokande~\citep{Super-Kamiokande:2015qek} (round marker) and IceCube~\citep{IceCube:2015mgt, IceCube:2016umi} (triangle marker) for electron neutrinos (grey) and muon neutrinos (black). Predicted combined muon and electron neutrino atmospheric flux from Ref.~\citep{Honda:2011nf} (solid grey line) is also displayed.
[Right] DM annihilation cross-sections for which the DS neutrino flux peak exceeds that of the atmospheric background for a selection of $f_\textrm{SMSD}$ and $f_{\nu} = 1/3$. Line colors indicate where Super-Kamiokande (green) and IceCube (magenta) have leading sensitivity. Steep DS collapse model (thick lines) and gradual DS collapse model (thin lines), consistent with JWST candidate events, are shown. We assume DSs grow to 1\% of their host halo mass in $M_h = 10^6-10^9~M_\odot$ halos, to an upper mass limit of $M_{\rm DS} = 10^6 M_\odot$. Overlaid (shaded regions) are existing bounds from indirect DM detection searches in IceCube~\citep{IceCube:2023ies}, Super-Kamiokande~\citep{Super-Kamiokande:2020sgt}, and ANTARES~\citep{Gozzini:2021iwl} derived assuming $f_\nu=1$. The thermal relic DM annihilation cross-section~\citep{relic_cross_sec} (gray dashed line) is displayed for reference.}
\label{fig:DS_sens}
\end{figure*}

DS parameters consistent with JWST observations can be identified from the number of supermassive DSs observable in JWST at a particular redshift expressed as
\begin{equation}\label{eq:observ} 
    \frac{dN_\textrm{\rm obs}}{dz} =  \frac{\theta^2}{4\pi}\int   dM_{\rm DS} f_{\rm sur}(z, M_{\rm DS})    f_{\rm obs}(z, M_{\rm DS})  \frac{dV_c}{dz} \frac{dn_{\rm DS}}{dM_h} (M_h, z(t-\tau))~,
\end{equation}
where $\theta^2=26.4$~arcmin$^2$ is the JWST coverage~\citep{Rieke:2023tks, 2023arXiv230602465E, deugenio2024}, $f_{\rm sur}$ and $f_{\rm obs}$ denote the DS fraction that survives to redshift $z$ and the likelihood that JWST observes a DS of mass $M_{\rm DS}$ at redshift $z$ respectively. Eq.~\eqref{eq:observ} also includes the differential volume element $dV_c/dz$, with $V_c = (  4\pi/3) D_M^3$ the comoving volume of the Universe where~\citep{Kolb:1990vq}
\begin{equation}
    D_M = \left(\frac{2c}{H_0}\right) \frac{2-\Omega_M(1-z)-(2-\Omega_M)\sqrt{1+z\Omega_M}}{\Omega_M^2(1+z)}
\end{equation}
for a flat Universe without dark energy that is a good approximation at redshifts $z\gtrsim 10$ we consider.  The total number of DSs observable with JWST is the integral of Eq.~\eqref{eq:observ} over redshifts. 

Here we consider $f_\textrm{obs} (z > 15) = 0$ and survival model $f_\textrm{sur}$ being a step-function at $z=15$, resulting in conservative $N_\textrm{obs}=0$ DS candidates for all redshifts. This also maximizes possible DS neutrino flux consistent with JWST observations. 
In Appendix~\ref{app:limits} we present a model of gradual DS collapse, showing that it can account for $N_\textrm{obs} \geq 1$ JWST DS candidates. We observe a general feature that extended history of DS collapse, observations such as by JWST can constrain the entire model by probing the lower redshift surviving tail of the DS population. Hence, the gradual collapse model results in a smaller neutrino flux than the step-function model.  Our analysis can be also readily adopted to other DS collapse histories.

\section{Dark star neutrino signals} Supermassive DSs shining at high luminosities can emit significant neutrino fluxes, contributing to a diffuse background analogous to the diffuse supernova neutrino background (DSNB) in the later Universe.
Unlike the DSNB produced by transient events with neutrino energies determined by supernova temperatures, DSs emit neutrinos over prolonged time with neutrino energies determined by DM.  

The electromagnetic comoving luminosity density of the total DS population as a function of redshift and integrated by DS age-distribution is 
\begin{equation}\label{eq:pop_lum} 
    L_\textrm{EM}(z) =   \int_0^{t(z)} d\tau \int_{M_{h,\textrm{min}}}^{M_{h,\textrm{max}}}    dM_h L_{\rm DS}(M_h, \tau)      f_{\rm sur}(z)  \frac{d^2n_{\rm DS}}{dM_hdt} (M_h, z(t-\tau))~,
\end{equation}
where evaluating $d^2n_{\rm DS}/dM_hdt$ by differentiating Eq.~\eqref{eq:mass_func} and evaluating at $z(t-\tau)$ gives the DS age $\tau$ distribution at cosmological time $t$ formed in halos of mass $M_h$. Here, we do not include contributions from DM annihilation outside of DSs that is also enhanced by the adiabatic contraction, making our estimates conservative. 

The resulting DS diffuse neutrino flux 
\begin{equation}\label{eq:flux} 
    \frac{d\phi}{dE_\nu} =  \int_{z_{\rm lim}}^\infty dz \bigg[(1+z) \frac{dN}{dE_\nu}(E_\nu (1+z))   \left(\frac{f_\nu }{(1-f_\nu) \langle E_\nu\rangle} L_\textrm{EM} (z)\right)\left|c\frac{dt}{dz}\right|\bigg]
\end{equation}
where $\langle E_{\nu} \rangle$ is the average neutrino energy from DM annihilation and depends on considered channel,
$dN/dE_\nu$ is a neutrino emission spectrum from DM annihilation and spectrum factor $(1+z)$ accounts for compression of the energy scale. In Ref.~\citep{Freese_2016} $f_{\nu} = 1/3$ was assumed.  For DM annihilation channels we consider neutrino spectra computed by Ref.~\citep{Marco_Cirelli_2011, Paolo_Ciafaloni_2011}.
Here, $f_\nu/(1-f_\nu)\langle E_\nu\rangle$ converts the electromagnetic luminosity to neutrino number flux. We integrate from a minimum redshift where JWST becomes unable to observe the most massive DSs in our population $z_{\rm lim} = 15$. 
In the Appendix we show that our generalized formalism for extended source emission can also account for fluxes like DSNB~\citep{Beacom:2010kk} and compare to additional literature on DM annihilation~\citep{Arguelles:2019ouk, Prada:2011jf, Lopez-Honorez:2013cua, Yuan:2011yb, Schleicher:2008aa}.

\section{Constraints from Neutrino Experiments} In Fig.~\ref{fig:DS_sens} we display resulting DS neutrino flux at present for different DM annihilation channels and masses computed with Eq.~\eqref{eq:flux}, together with atmospheric neutrino background. We focus on signals when $E_{\nu} \gtrsim 200$~MeV. At lower energies contributions from solar, DSNB and reactor neutrinos also appear.

The DS neutrino flux has cut-off energy spectrum within factor of few of $E_{\nu} \sim m_{\chi}/(z_{\rm lim} + 1) \sim m_{\chi}/16$, set by the DM mass and considered annihilation channel as well as redshift effects. Hence, such neutrino flux could be distinguished with  atmospheric neutrino background including by spectral shape, depending on uncertainties. Further, at higher energies muon neutrinos dominate atmospheric flux~\citep{IceCube:2015mgt}, while the DS neutrino flux can contain all flavors. For lower DM masses we find Super-Kamiokande~\citep{Super-Kamiokande:2015qek} data can place the strongest bounds, while data from IceCube~\citep{IceCube:2015mgt, IceCube:2016umi} for $m_{\chi}\gtrsim 10$~TeV.

We show DM parameters for which neutrino flux peak from DSs is estimated to exceed the atmospheric neutrino background in  Fig.~\ref{fig:DS_sens} (right panel) when DM annihilation has has $f_{\nu} = 1/3$ to $\nu\bar{\nu}$ for energies $E_{\nu} \gtrsim 200$~MeV following Super-Kamiokande~\citep{Super-Kamiokande:2015qek} measurements,
well fit by predictions~\citep{Honda:2011nf},  and at higher energies IceCube~\citep{IceCube:2015mgt, IceCube:2016umi} data. For reference, we overlay
existing indirect detection constraints on DM annihilation. These estimated limits are seen to improve at higher DM masses due to different power law scaling between atmospheric flux and DS flux and at lower masses where the atmospheric flux flattens. Further, sensitivity at smaller DM masses increases also due to increased luminosity as seen from Eq.~\eqref{eq:Lum_approx}.
These results can be further refined with detailed statistical analysis that we leave for future work. We estimated that a $\chi^2$ analysis considering available data and uncertainties on detected neutrino energies  from Super-Kamiokande~\citep{Super-Kamiokande:2015qek} can be expected to yield results within $\sim 1-2$ orders of magnitudes of our present simplified treatment.

At lighter masses the DS neutrino events become degenerate with those of DSNB, although with distinct spectra. 
At larger DM cross-sections, the electromagnetic emission from DSs is also enhanced and can affect other constraints. 
In Appendix~\ref{app:results} we discuss DM annihilation to $W^+W^-$, $b\bar{b}$, and $\mu^+\mu^-$~\citep{HAWC:2023owv, Fermi-LAT:2015att, AMS:2014xys, AMS:2014bun, HESS:2016mib}. While we focus on neutrino emission, we note potential complementarity with other signals such as gamma rays~\citep{CTAO:2024wvb} that depend on DM interactions and regarding which we remain agnostic.
Upcoming large scale neutrino telescopes such as the 187-kton water Cherenkov detector Hyper-Kamiokande~\citep{Hyper-Kamiokande:2018ofw}, 40 kton liquid argon DUNE~\citep{DUNE:2020lwj, DUNE:2015lol} and 20 kton liquid scintillator JUNO~\citep{JUNO:2021vlw} can probe DSs fueled by $\sim$GeV–TeV scale DM with enhanced sensitivity.  

\section{Conclusions} DS can constitute compelling candidates for the earliest stars and provide seeds for SMBHs. We demonstrated that massive $\gtrsim 10^5 M_{\odot}$ DSs can imprint observable diffuse neutrino flux with energies that can extend to hundreds of MeVs in existing experiments.
Considering DS neutrino emission we have established the first constraints on DS abundance and DM annihilation cross-sections that fuel them using Super-Kamiokande and IceCube data. Our results are less sensitive to assumptions about DS population and masses compared to direct searches, offering a complementary approach to probing them at high redshifts.  This opens a new route into probing SMBH progenitors. Upcoming neutrino detectors such as Hyper-Kamiokande, DUNE and JUNO Future will be able to probe DS properties with enhanced sensitivity.

~\newline
\begin{acknowledgments}
    We thank John Beacom, Katherine Freese, Tracy Slatyer, Anna Suliga, and Tien-Tien Yu for discussions. V.T. acknowledges support by the World Premier International Research Center Initiative (WPI), MEXT, Japan and JSPS KAKENHI grant No. 23K13109. This work was performed in part at the Aspen Center for Physics, which is supported by the National Science Foundation grant PHY-2210452.
\end{acknowledgments} 

\appendix

\section{Dark Star Model and Evolution}\label{app:polytrope}

We analytically investigate DS evolution by considering a polytrope model with hydrostatic and thermal equilibrium. 
This is expected to be in approximate agreement with detailed numerical evaluations of DS evolution with MESA 1D stellar evolution within a factor of few~\citep{Rindler-Daller:2014uja}.

DSs are powered by DM heating instead of nuclear fusion in their cores and are of lower density and temperature than main sequence stars. With their equation of state typically dominated by non-relativistic gas pressure, DS stellar density profiles in early stages can be modeled as a polytrope with index $n = 3/2$~\citep{Freese_2008}, describing systems supported primarily by thermal pressure with contributions from DM annihilation. In Ref.~\citep{Spolyar:2009nt} later stages of DS evolution were modeled as $n = 3$ polytropes, considering dominance of radiation pressure. 
We confirmed that $n=3$ polytropes result in larger DS volume and lower temperatures than their $n=3/2$ counterparts, with resulting luminosity differing within a factor of few. Here, we use $n = 3/2$ throughout.
The equation of state relating pressure $P$ and density $\rho$ is
\begin{equation}\label{eq:poly}
    P = K \rho^{1+1/n}~,
\end{equation}
where $n$ is the polytrope index and constant $K$ is set by the stellar boundary conditions. We focus on $n = 3/2$ polytrope description of DSs.

For mass $M_r$
contained within a radius $r$  
\begin{equation}
    M_r = \int_0^r dr' 4\pi r'^2\rho(r')
\end{equation}
the hydrostatic equilibrium is
\begin{equation}\label{eq:hydro_eq}
    \frac{dP}{dr} = -\rho\frac{GM_r}{r^2}~.
\end{equation}
Combining with Eq.~\eqref{eq:poly}, this gives the well known Lane-Emden equation
\begin{equation}\label{eq:lane_em}
    \frac{1}{x^2}\frac{d}{dx}\left(x^2\frac{d\theta}{dx}\right) = -\theta^n~,
\end{equation}

\noindent in terms of the dimensionless variables $x = r/\alpha$, $\theta^n = \rho/\rho_c$ where $\alpha = (n+1)K/4\pi G \rho_c^{1-1/n}$ and $\rho_c$ is the density at the center of the star. This is subject to the boundary conditions $\theta(0)=1$, $d\theta/dx(0) = \theta'(0)=0$, and $\theta(x_1)=0$ where $x_1$ corresponds to the surface of the star. For $n=3/2$, one finds $x_1\simeq3.65$.
For a particular stellar radius $R$ and mass $M$ the constant $K$ from Eq.~\eqref{eq:poly} is determined by 
\begin{equation}\label{eq:K}
    K = \frac{1}{n}\left[\left(\frac{R}{x_1}\right)^{-1+3/n}\left(\frac{GM}{-x_1^2\theta'(x_1)}\right)^{1-1/n}(4\pi G)^{1/n}\right]~.
\end{equation}
Here, we use the density contrast $D_n = \rho_c/\bar{\rho}$, which evaluates to $D_{3/2} = 5.99$ for $n = 3/2$, in order to set $\rho_c$.

\begin{figure}[t]
    \centering
\includegraphics[width=0.45\linewidth]{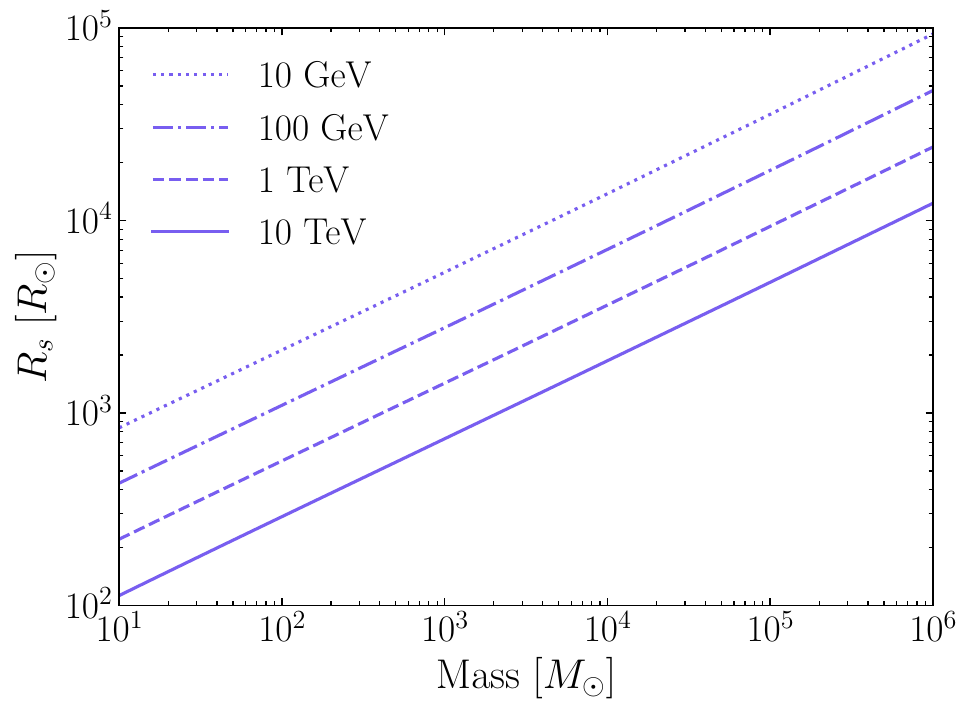}
    \caption{DS photosphere radius for DM annihilation channel $\nu\bar{\nu}$ for different masses considering that the fraction of DM mass $m_{\chi}$ converted to neutrino energy is $f_\nu=1/3$.}
    \label{fig:DS_rad}
\end{figure}

We can then compute the photosphere radius $R_s$. The temperature is governed by the equation of state as
\begin{equation}\label{eq:EoS}
    P(r) = \frac{\rho k_B T(r)}{\bar{m}} + \frac{4}{3}\frac{\sigma_B}{c}T(r)^4~,
\end{equation}
where $\sigma_B$ the Stefan-Boltzmann constant, $k_B$ is the Boltzmann constant, and $\bar{m} = m_u(2X+3Y/4)^{-1} \simeq 0.588 m_u$ is the mean atomic weight in terms of mean atomic mass unit $m_u\simeq 0.931$ GeV with H and He mass fractions $X=0.76$ and $Y=0.24$, respectively, taken from Big Bang nucleosynthesis~\citep{Planck:2015fie}. 
The resulting photosphere radius can be obtained by solving
\begin{equation}
P \kappa\left(\rho_{R_s},T_{R_s}\right) = \dfrac{2}{3}g(R_s) ~,
\end{equation} 
where $\kappa$ is the opacity taken from numerically computed OPAL astrophysical opacity tables~\citep{1996ApJ...464..943I} and $g(r)=GM_r/r^2$ is the gravitational acceleration. Here, $\rho(r)$ is determined from the Lane-Emden equation \eqref{eq:lane_em}, $P$ from the polytrope equation \eqref{eq:poly}, and $T$ from the equation of state \eqref{eq:EoS}. 
The surface luminosity is then determined from the obtained photosphere radius and effective temperature considered to be $T(R_s)$
\begin{equation}\label{eq:surface_lum}
    L_{\rm surf} = 4\pi\sigma_BR_s^2T(R_s)^4 ~.
\end{equation}

We consider three distinct DS sources of energy playing a relevant role: (A) nuclear fusion, (B) gravitational energy and (C) DM heating via annihilation. Besides contributions from initial DM density, DM heating can benefit from the capture of additional DM from scattering with baryons  in the star. We do not consider this contribution as it relies on additional assumptions about the DM-baryon interactions. Once DM is depleted, the star contracts and transitions to hydrogen burning. Prior to this, DM contributions dominate due to the lower core densities and pressures that suppress nuclear fusion, and the efficiency of DM annihilation. As DM annihilation products are converted into heat and heating the star more efficiently than fusion, neutrinos escape~\citep{Spolyar:2007qv}.
Therefore, we have 
\begin{equation}\label{eq:Lums}
    L_{\rm tot} = L_{\rm fus} + L_{\rm grav} + L_{\rm DM} \simeq L_{\rm DM}~.
\end{equation}

The DM luminosity can be found from
\begin{equation}\label{eq:L_dm}
    L_{\rm DM} \simeq (1-f_\nu)\int_0^R dr 4\pi r^2   \dfrac{\langle \sigma v \rangle \rho_\chi^2}{m_\chi}~.
\end{equation}
Note that eq. \eqref{eq:L_dm} is enhanced by a factor of two for Majorana particle DM.
From Navarro-Frenk-White (NFW) DM profile~\citep{Navarro:1996gj} simulations, after adiabatic contraction the DM density at the outer edge of baryonic core is found to be~\citep{Spolyar:2007qv}  
\begin{equation}\label{eq:dm_dens}
    \rho_\chi \simeq 1.7\times 10^{11}~{\rm GeV}{\rm cm}^{-3}\Big(\dfrac{n_b}{10^{13}~\textrm{cm}^{-3}}\Big)^{0.81}~,
\end{equation}
\noindent where $n_b$ %
is the baryon number density. We consider $n_b = 10^{13}~\textrm{cm}^{-3}$ at the initial DS formation, as in Ref.~\citep{Spolyar:2007qv}. We then iteratively solve the polytrope equations. As a DS
evolves, $n_b$ changes accordingly. Outside the core, the DM is found to scale with $r^{-1.9}$. We adopt this as the representative DM density.

\begin{figure}[t]
    \centering
    \includegraphics[width=0.45\linewidth]{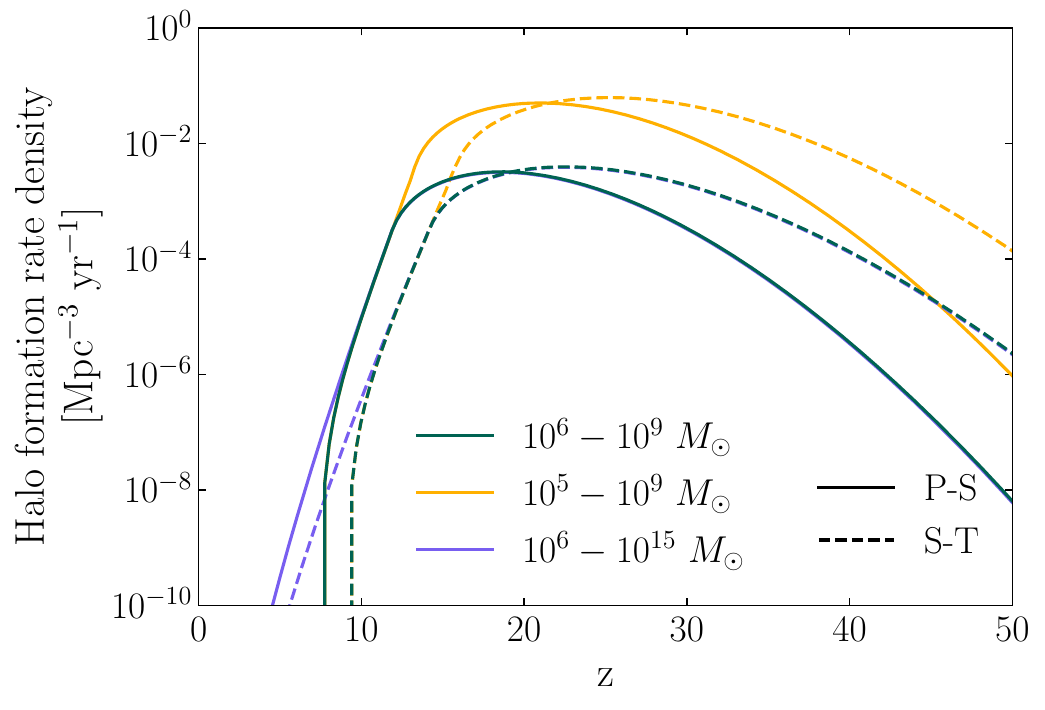}
    \caption{Halo formation rate density $d^2 n/d M_h dt$ integrated over halo mass range. Integrated halo formation rate density we consider as reference (green line), results extended to lighter halos (yellow line), results extended to more massive halos (lavender line) are shown.
    Press-Schechter (solid lines, P-S) and Sheth-Tormen (dashed lines, S-T) halo formation are indicated.}
    \label{fig:halo_MF}
\end{figure}

In thermal equilibrium, the luminosity of Eq.~\eqref{eq:L_dm} must balance the surface luminosity of Eq.~\eqref{eq:surface_lum}, with
\begin{equation} \label{eq:theq}
    L_{\rm DM, eq.} = L_{\rm surf, eq}~.
\end{equation}
In Fig.~\ref{fig:DS_rad} we display
the equilibrium radius as a function of time for DM annihilating to $\nu\bar{\nu}$ in a DS that reaches $10^6~M_\odot$ after prolonged accretion at $\dot{M} = 0.1~M_\odot$ yr$^{-1}$. 
To compute this, we first consider an initial DS radius value and find its photosphere radius. We then compare the surface and DM luminosities. After this, we iteratively adjust our initial radius as needed until Eq.~\eqref{eq:theq} is satisfied. This procedure yields resulting DS luminosity and radius as well as temperature, density, and pressure profiles for a given DS mass and DM particle mass.  We have verified that for other DM decay channels results are similar. 

Incorporating numerical simulations, analysis of Ref.~\citep{Freese_2010} revealed that centrophilic orbits~\citep{valluri_orbital_2010}, which repeatedly pass through the star, can sustain the DM fuel supply for 
$\gtrsim1$~Gyr. This extended fueling allows to support supermassive DSs, with their masses reaching $\gtrsim 10^6~M_\odot$ and luminosities of $10^9 - 10^{10}~L_\odot$. 
In our study we also neglect the possibility of DM capture via scattering off baryons in DSs, which could extend their lifetime after the ambient DM density decreases.
The effectiveness of this mechanism depends on the DM interactions with SM. The impact of DM capture on dark stars was initially studied in~Ref.~\citep{Freese_2008_capture} and further refined in~Ref.~\citep{Freese_2016}, where it was shown that stars fueled by captured DM are typically few times hotter and have radii smaller by an order of magnitude compared to those powered solely by adiabatic contraction.

Our results for supermassive DSs in Fig.~\ref{fig:DS_evo} give luminosities that are approximately consistent with discussion of Ref.~\citep{Rindler-Daller:2014uja} indicating that luminosities obtained from polytropic model are expected to be suppressed by a factor of few compared to results from detailed stellar evolution simulations. We find that the DS luminosity can be well approximated by Eq.~\eqref{eq:Lum_approx}.


\section{Dark Star Population and  Collapse Rate}\label{app:limits}

The halo density formation rate per halo mass $d^2 n /d M_h dt$ can be obtained by differentiating Eq.~\eqref{eq:HFR} with respect to time. 
Integrated contributions over halo masses of the halo density formation rate in the total DS luminosity as described by Eq.~\eqref{eq:pop_lum} 
depend on the considered halo mass limits $M_{h,\textrm{min}}$ and $M_{h,\textrm{max}}$. In Fig.~\ref{fig:halo_MF} we illustrate the effect of different choices of the halo mass limits on the integral of $d^2n/dM_hdt$ with respect to $M_h$. Increasing $M_{h,\textrm{max}}$ does not have an appreciable effect at $z_{\rm lim}$ since 
massive halos become increasingly rare at higher redshifts. On the other hand, decreasing $M_{h,\textrm{min}}$ substantially increases the halo density formation rate due to larger contributing population of smaller halos. These lighter halos are less luminous and thus result in a smaller enhancement on the flux despite being numerous. 

 \begin{figure}[t]
    \centering
    \includegraphics[width=.45\linewidth]{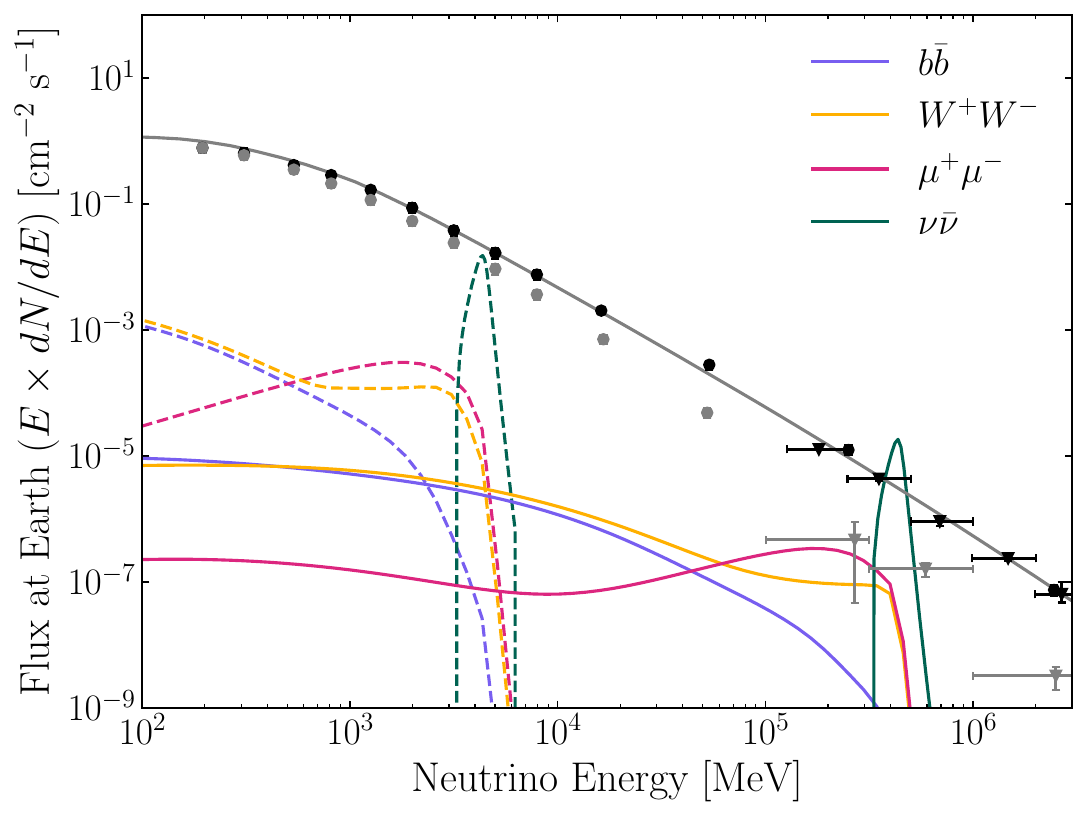}
    \caption{Same as Fig.~\ref{fig:DS_sens} considering gradual DS collapse model with $f_\textrm{sur}$ given by Eq.~\eqref{eq:gradcol}.}
    \label{fig:collapse_models}
\end{figure}

As discussed in the main text, DSs can potentially be detected by sensitive telescopes such as JWST. For $10^{6}~M_\odot$ DSs, with luminosities of $10^{10}~L_\odot$, we consider that JWST can observe all such objects that survive beyond $z=15$. 
Three DS candidates have been identified considering JWST JADES survey~\citep{Ilie:2023zfv}, which initially covered approximately $\sim 26.4~\textrm{arcmin}^2$~\citep{2023arXiv230602465E, Rieke:2023tks}. Since the identification of these candidates, the JADES survey has expanded its observational area to $\sim 56$~arcmin$^2$~\citep{deugenio2024}.  
We assume JWST can efficiently identify such DSs. Then, to remain consistent with these observations, the majority of
$\sim10^6~M_\odot$ DSs should collapse before $z=15$. 

In our analysis, we assume the simplest possible history of DS collapse and   model the surviving population with a step function in redshift
\begin{equation} \label{eq:heaviside}
    f_\textrm{sur} = 1 - \Theta(z_\textrm{lim} - z)
\end{equation}
where $\Theta$ is the Heaviside step function and $z_{\rm lim}=15$ is motivated by JWST sensitivity. Other DS collapse histories can be readily implemented within our analysis framework. The actual DS lifetime evolution is complex and requires dedicated simulations beyond the scope of our work. 
we also consider a more gradual collapse scenario, with a DS survival fraction parameterized by
\begin{equation} \label{eq:gradcol}
    f_{\rm sur}=\dfrac{1}{2}\Big[1+\tanh\left(z-z_0\right)\Big]~,
\end{equation}
where $z_0$ is a redshift parameter.
In the framework of Eq.~\eqref{eq:gradcol}, a fraction of the population survives to later times. Observations by JWST allow to constrain the surviving DS fraction and combination $f_\textrm{SMDS} f_{\rm sur}$. 
Choosing $f_\textrm{SMDS}=10^{-2}$ as a reference values, consideration of $z_0 \simeq 22$ allows for three candidate objects in the 26.4 arcmin$^2$ JADES field of view as claimed by analysis of Ref.~\citep{Ilie:2023zfv}.  Additional data from JWST will expand the angular coverage $\theta^2$. If no additional DS candidates are found, the resulting constraints on the population of luminous DSs will become more strict. Specifically, the bounds on the number of detectable DSs at high luminosities can be improved, with results scaling as $\propto N/\theta^2$, see Eq.~\eqref{eq:observ}. Our approach establishes a direct connection between DS diffuse neutrino emission and the DS candidates of JWST.

In Fig.~\ref{fig:collapse_models}, we display
the predicted DS neutrino flux considering gradual DS collapse with $f_{\rm sur}$ following Eq.~\eqref{eq:gradcol}
and compare with 
Fig.~\ref{fig:DS_sens} based on step function rapid DS collapse model of Eq.~\eqref{eq:heaviside}. The gradual DS collapse model gives neutrino flux predictions suppressed by around an order of magnitude compared to the step-function model. This originates from constraints on the low-redshift tail of the DS population that also impact DS population at higher redshifts. 


\section{Generalized Diffuse Neutrino Flux and Comparison with DSNB}\label{app:DSNBflux}

Our general formalism of Eq.~\eqref{eq:flux}, and considering Eq.~\eqref{eq:pop_lum} as the comoving luminosity density of population of sources, captures extended source emission with variety of source lifetimes. This can readily reduce to conventional formalism for transient sources~\citep{Beacom:2010kk}. In particular, we can consider the DSNB with evolving redshift-dependent core-collapse supernova rate $R_{\rm SN}(z)$.

Let us consider DS luminosity as a Dirac delta function 
\begin{equation}\label{eq:delta_func}
    L_{\delta} = L_0 \delta(\tau) ~,
\end{equation}
with $L_0 = \int d\tau L_{DS}(\tau)$ such that the total integrated DS luminosity is identical to that of prolonged emission. Note that Eq.~\eqref{eq:delta_func} also implies all DS emission occurs at formation, with $\tau=0$, allowing to neglect $f_\textrm{sur}$. 
With this luminosity, Eq. \eqref{eq:pop_lum} becomes 
\begin{align}\label{eq:lum_delta}
    L_{\rm EM}(z) =&~ \int_0^{t(z)}d\tau\int_{M_{min}}^{M_{max}} dM_h L_0 \delta(\tau) \frac{d^2n_{\rm DS}}{dM_hdt}\Big(z(t-\tau),M_h\Big) \notag\\
    =&~ L_0 \int_{M_{min}}^{M_{max}} dM_h \frac{d^2n_{\rm DS}}{dM_hdt}\Big(z(t), M_h\Big) = L_0 R(z)~.
\end{align}
where in Eq. \eqref{eq:lum_delta} we have identified $R(z)$ as the rate density of DS formation that is analogous to supernova rate $R_{\rm SN}(z)$ in computation of DSNB~\citep{Beacom:2010kk}. 

\begin{figure}[t]
    \centering
    \includegraphics[width=0.45\linewidth]{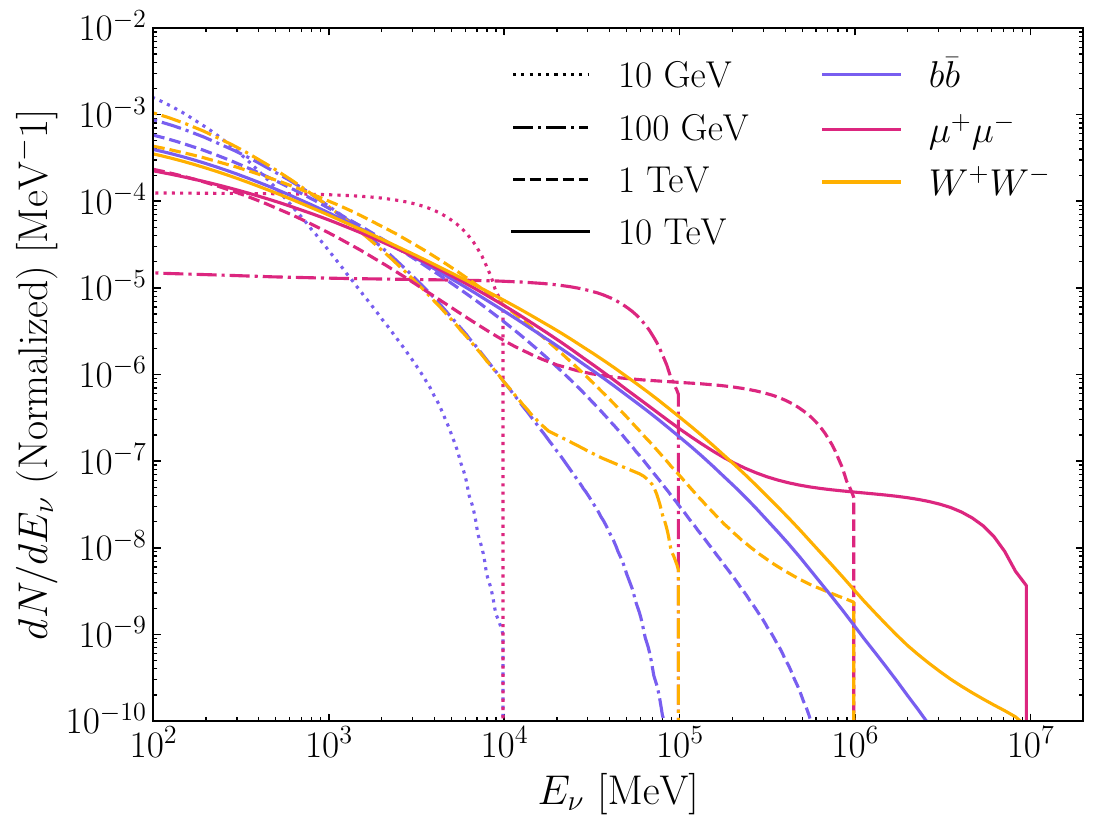}
    \caption{Differential neutrino spectra normalized to one neutrino for a range of DM masses from $m_{\chi} = 10$~GeV to 10~TeV for different DM annihilation channels. Spectra are taken from Ref.~\citep{Marco_Cirelli_2011, Paolo_Ciafaloni_2011}}
    \label{fig:nu_spec}
\end{figure}

From this computation, our Eq. \eqref{eq:flux}  becomes Eq. \eqref{eq:delta_flux}, where $L_0 f_\nu/(1-f_\nu)\langle E_\nu\rangle$ is the number flux of neutrinos emitted per DS and spectrum $dN/dE$ is normalized. Hence, $(dN/dE_{\nu}L_0 f_\nu/(1-f_\nu)\langle E_\nu\rangle$
is the neutrino emission from a single DS analogous to the DSNB emission spectrum $\varphi$ in Ref.~\citep{Beacom:2010kk} in units of neutrinos per unit energy. Under theses considerations, Eq. \eqref{eq:delta_flux} can be matched with DSNB flux computation of Ref.~\citep{Beacom:2010kk}  
\begin{equation}\label{eq:delta_flux}
    \frac{d\phi}{dE_\nu} = \int dz \left[(1+z)\frac{dN}{dE_\nu}\Big(E_\nu(1+z)\Big)\frac{f_\nu/(1-f_\nu)}{\langle E_\nu\rangle}L_0\right]\left[ R(z)\right]\left|c\frac{dt}{dz}\right|~.
\end{equation}
In Fig.~\ref{fig:nu_spec}
we show the neutrino spectra $dN/dE$ obtained from Ref.~\citep{Marco_Cirelli_2011} and list a selection of neutrino fractions in Tab.~\ref{tab:f_nu}.

\begin{figure}
    \centering
    \includegraphics[width=0.45\linewidth]{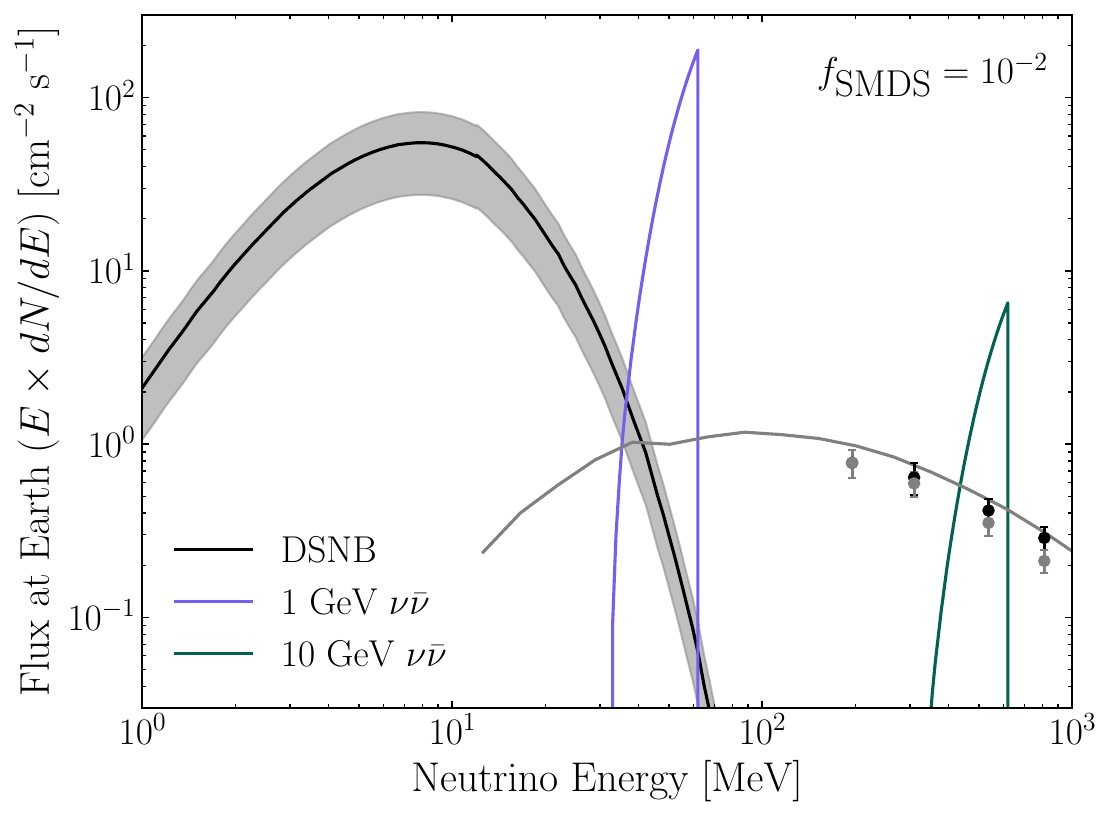}
    \caption{Comparison of the DSNB and DS neutrino fluxes for 1 GeV and 10 GeV DM annihilating to $\nu\bar{\nu}$.}
    \label{fig:dsnb}
\end{figure}

It is insightful to compare the predicted DSNB flux with DS contributions. The DS number density can be estimated  using  
\begin{equation}
n_\textrm{DS} = f_\textrm{SMDS} \int_{M_\textrm{h,min}}^{M_\textrm{h,max}} \frac{dn_h}{dM} dM,
\end{equation}
where we take $f_\textrm{SMDS} = 10^{-2}$ and $dn_h/dM$ as given in Eq.~\eqref{eq:HFR}. This yields approximately $\sim 4\times10^3$~Mpc$^{-3}$ DS number density at $z_\textrm{DS} = 15$. To compare with the DSNB flux originating from supernovae contributions around $z_\textrm{DSNB} \sim 1$, we can apply a rescaling factor of $(1+z_\textrm{DSNB}^3 / (1+z_\textrm{DS}^3$, obtaining an estimated DS number density of $\sim 8$~Mpc$^{-3}$ at $z = 1$.

Let us now consider that DS are powered by the annihilation of 10 GeV DM into neutrino-antineutrino pairs  $\nu\bar{\nu}$. Each such object radiates at approximately $\sim 10^{10} L_\odot$ for $\sim 10^8$ years, producing a total of $\sim 10^{61}$ neutrinos.
The core-collapse supernova rate can be roughly estimated as $\sim 10^{-3}$~Mpc$^{-3}$~yr$^{-1}$ and considering contributions primarily over redshift range of $z \simeq 1.5$ to $z \simeq 0.5$, spanning $\sim 10^9$ years. This corresponds to a total of $\sim 10^6$~Mpc$^{-3}$ supernovae, with each supernova emitting $\sim 10^{58}$ neutrinos~\citep{Beacom:2010kk}. Consequently, supernovae outnumber DS by a factor of $\sim 10^5$, while an individual DS produces $
\sim10^3$ times more neutrinos than a single supernova. Combining these effects, we observe that this leads to a expectation that the DSNB flux should exceed the contributions from DSs powered by annihilation of 10 GeV DM to neutrinos by roughly two orders of magnitude $\sim 10^2$, although the two fluxes peak at distinct energies. This is confirmed in Fig.~\ref{fig:dsnb}, where the integrated DSNB flux is found to be a factor of $\sim 90$ larger than that from DSs fueled by 10 GeV DM annihilation. 

Additionally, we can compare DSNB and DS flux contributions in case the mass of annihilating DM is 1 GeV. The total number of neutrinos from a given DS population scales as $\sim L_\textrm{DS}/m_{\chi}$. From Eq.~\eqref{eq:Lum_approx}, reducing $m_{\chi}$ by a factor of 10 increases $L_\textrm{DS}$ by a factor of $\sim 3$, leading to an increase in the DS neutrino flux by a factor of $\sim 30$. Hence, the DS neutrino flux is lower than DSNB flux by a factor of $\sim 3$. This is in agreement with what we find in Fig.~\ref{fig:dsnb} after integration, where we also observe the expected shift to lower neutrino energies of peaks for smaller DM mass annihilation.

\begin{table}[t]
    \centering
    \begin{tabular}{c|c c | c c | c c | c c}
    \hline
    \hline
      Annihilation         & \multicolumn{2}{c|}{$m_{\chi}$ = 10~GeV} & \multicolumn{2}{c|}{$m_{\chi}$ = 100~GeV} & \multicolumn{2}{c|}{$m_{\chi}$ = 10~TeV} & \multicolumn{2}{c}{$m_{\chi}$ = 100~TeV} \\ channel 
                       & $f_\nu$ & $\langle E_\nu \rangle$ (GeV) & $f_\nu$ & $\langle E_\nu \rangle$ (GeV) & $f_\nu$ & $\langle E_\nu \rangle$ (GeV) & $f_\nu$ & $\langle E_\nu \rangle$ (GeV) \\ \hline
        $b\bar{b}$     & 0.47  & 0.224  & 0.45  & 0.944   & 0.45  & 4.19   & 0.45  & 19.8  \\
        $\mu^+\mu^-$   & 0.63  & 3.13  & 0.62  & 31.0  & 0.62  & 206  & 0.61  & 1113  \\
        $W^+W^-$       & N/A  & N/A   & 0.50  & 1.28  & 0.52 & 9.32   & 0.50 & 46.3   \\
        $\nu\bar{\nu}$ & 1/3  & 10   & 1/3  & 100   & 1/3   & $10^3$  & 1/3 & $10^4$  \\
        \hline\hline
    \end{tabular}
    \caption{Fraction $f_\nu$ of DM mass $m_{\chi}$ emitted as neutrino energy different DM masses and annihilation channels. The $\nu\bar{\nu}$ channel only produces neutrinos and hence does not inherently allow powering DSs through heating when $f_\nu=1$, and we consider $f_\nu\sim 1/3$.}
    \label{tab:f_nu}
\end{table}

\begin{figure}[t]
    \centering
    \includegraphics[width=0.45\linewidth]{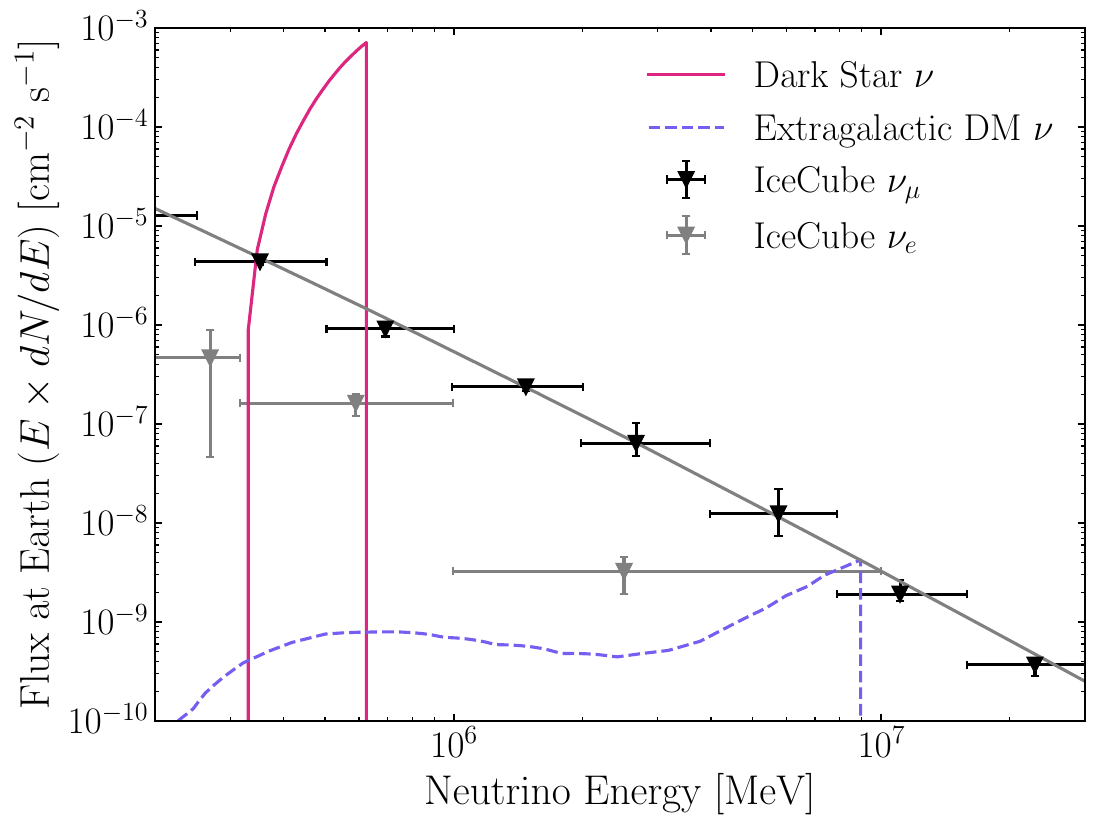}
    \caption{Extragalactic neutrino flux from 10 TeV DM annihilating in dark stars (pink/solid) assuming DS reach 1\% of their host halo mass which is in the range $M_h = 10^6 - 10^9~M_\odot$ compared to the expected neutrino flux from conventional dark matter halos (purple/dashed) from~\citep{Arguelles:2019ouk}.}
    \label{fig:extragalactic}
\end{figure}

\section{Comparison With Indirect Dark Matter Detection}\label{app:indirect-detection}

DM annihilation can produce a range of observable signatures, including   extragalactic diffuse gamma-ray background and the diffuse neutrino background. Here we examine how the emission from DSs associated with DM annihilation compares to signals in the context of indirect DM detection.

Since DM annihilation is follows $\propto \rho_{\rm DM}^2$ for DM density $\rho_{\rm DM}$, we define $f_{\rm DS}$ as the enhancement factor representing the overdensity of DM annihilation within DS compared to that in a typical galactic DM profile, which we assume follows the NFW~\citep{Navarro:1996gj} distribution
\begin{equation}\label{eq:f_ds}
    f_{\rm DS} = \dfrac{\int_0^{r_{\rm DS}} \rho_{\rm DS}^2(r)r^2dr}{\int_0^{r_\Delta} \rho_h^2(r)r^2dr}~.
\end{equation}
Here, $\rho_{\rm DS}$ is the DS density profile, $r_{\rm DS}$ is the DS radius and $\rho_{\rm NFW}$
is the NFW profile density 
  \begin{equation}
   \rho_h=\dfrac{4\rho_s}{(r/r_s)(1+r/r_s)^2}
  \end{equation} 
where $r_s$ is the characteristic   radius  and $\rho_s$ is the characteristic  density. 

We consider $r_\Delta$ to denote the outer radius of the halo such that
\begin{equation}
    M_h = \Delta \rho_c(z) \frac{4}{3}\pi r_\Delta^3~,
\end{equation}
where $\rho_c(z) \simeq \rho_c(\Omega_m(1+z)^3 + \Omega_\Lambda)$ is the critical density of the Universe with $\Omega_m$ and $\Omega_\Lambda$ denoting fractional contributions from matter and dark energy~\citep{Planck:2018vyg}. Hence, the density within $r_\Delta$ is $\Delta$ times the critical density of the Universe. Following Ref.~\citep{Prada:2011jf}, we consider overdensity threshold $\Delta = 200$.
Then, the halo concentration is found from $c_\Delta = r_\Delta/r_s$. We employ $c_{\Delta} (M_h, z)$ parametrization of Ref. \citep{Lopez-Honorez:2013cua} obtained from the fit of all available data from the MultiDark/BigBolshoi simulations~\citep{Prada:2011jf}, and which enters as
\begin{equation}
    \int_0^{r_\Delta}\rho_h^2(r)r^2dr = \tilde{g}(c_\Delta)\frac{M_h\Delta\rho_c(z)}{12\pi}
\end{equation}
where $\tilde{g}$ is given by
\begin{equation}
    \tilde{g}(c_\Delta) = \frac{c_\Delta^3[1-(1+c_\Delta)^{-3}]}{3[\ln(1+c_\Delta)-c_\Delta/(1+c_\Delta)]^2}~.
\end{equation}

Previously, analysis of DS gamma-ray background of Ref.~\citep{Yuan:2011yb} found that DM annihilation rate from DSs exceeds that of halos by enhancement of $f_{\rm DS}\sim10^3$. They considered DS density profiles of Ref.~\citep{Freese:2008hb} for $\sim 10-100 M_{\odot}$ DS forming in a $10^5-10^6~M_\odot$ halo.  Using of Ref.~\citep{Freese:2008hb} we have computed similar $f_{\rm DS}\sim10^3$ to that of Ref.~\citep{Yuan:2011yb}.
Importantly, those profiles correspond to very early phase in DS formation when baryonic matter has only contracted to a hydrogen density of $\sim10^{13}$cm$^{-3}$.

However, as can be seen from Ref.~\citep{Rindler-Daller:2014uja} 
for the supermassive DSs we consider, the baryon densities reach nearly $\sim 10^{20}$cm$^{-3}$.
For an increase in density from $10^{13}$cm$^{-3}$ to $10^{16}$cm$^{-3}$, considering the DS and halo mass is the same and DS mass is only $M_{\rm DS} \sim 10^{-4} M_h$, the enhancement factor significantly increases to $f_\textrm{DS} \sim10^6$. For supermassive DSs we study, as DSs further evolve and grow to $\sim1\%$ of the halo mass the enhancement factor increases to $f_\textrm{DS} \sim 10^8$. 
For a $M_h = 10^8~M_\odot$ halo we find $f_{\rm DS} = 4\times10^8$ for a DS of mass $M_{\rm DS} = 10^{6}$.

As we have demonstrated, supermassive DSs we explore benefit from greater enhancements $f_{\rm DS}$ compared to what was found in earlier studies ~\citep{Yuan:2011yb} that focused on less concentrated smaller and earlier DSs that compose a less significant fraction of the host halo mass.
We note that constraints from the extragalactic diffuse gamma-ray background have not been evaluated in such case in detail, which we leave for future work. We note that DS electromagnetic radiation has also been studied in the context of reionization~\citep{Schleicher:2008aa, Schleicher:2008gk}, however these analyses relied on a DS phases after DM depletion. 

\begin{figure*}[t]
    \centering
    \includegraphics[width=0.45\linewidth]{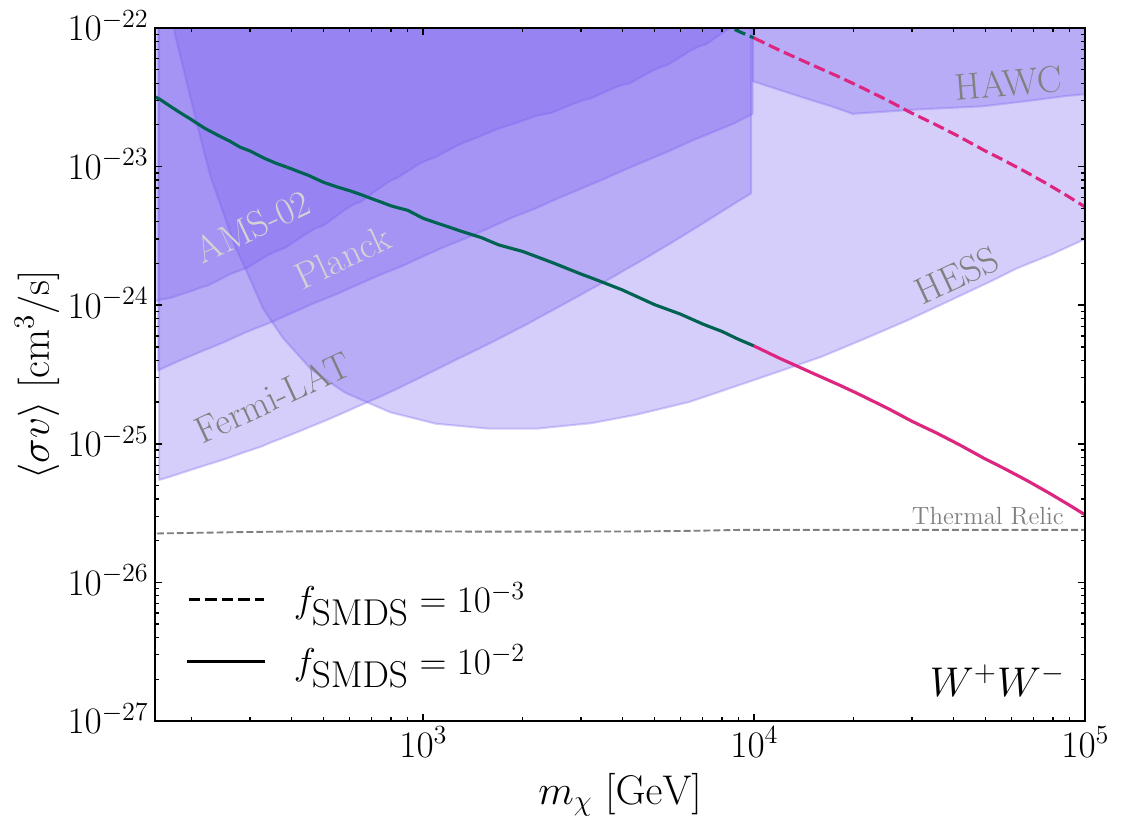}
        \includegraphics[width=0.45\linewidth]{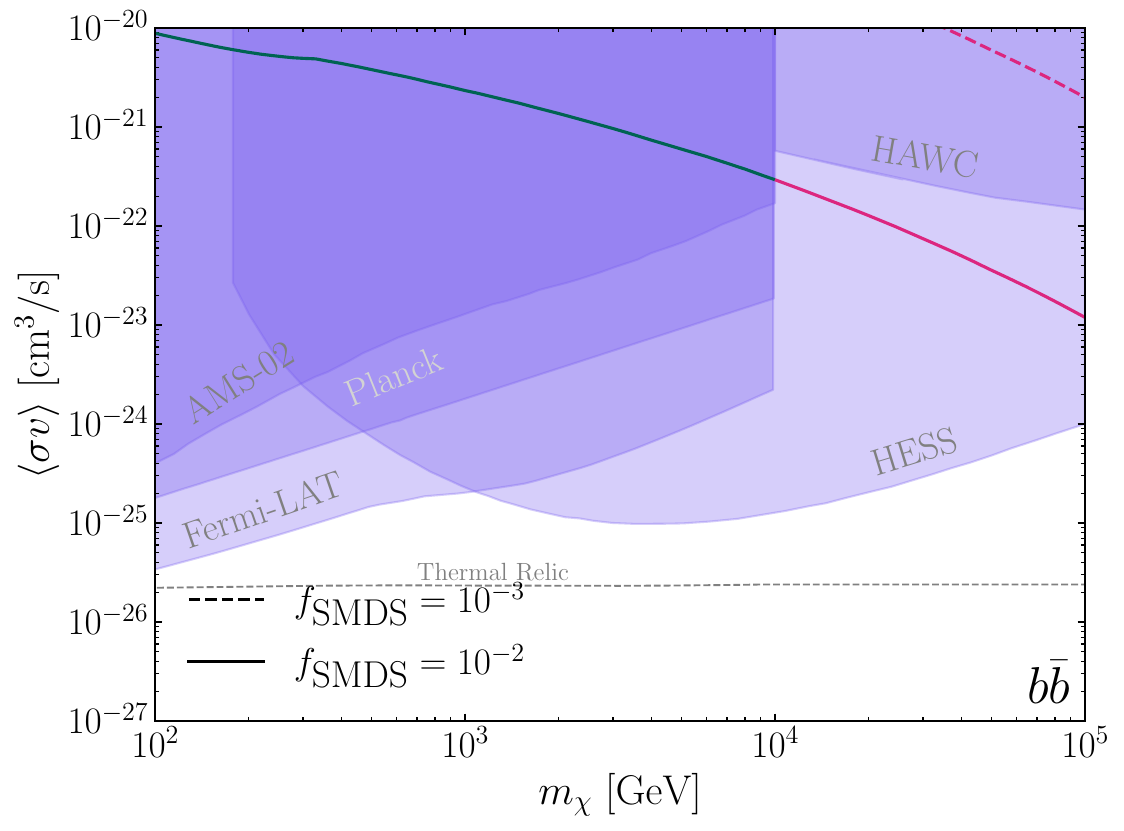}
        \includegraphics[width=0.45\linewidth]{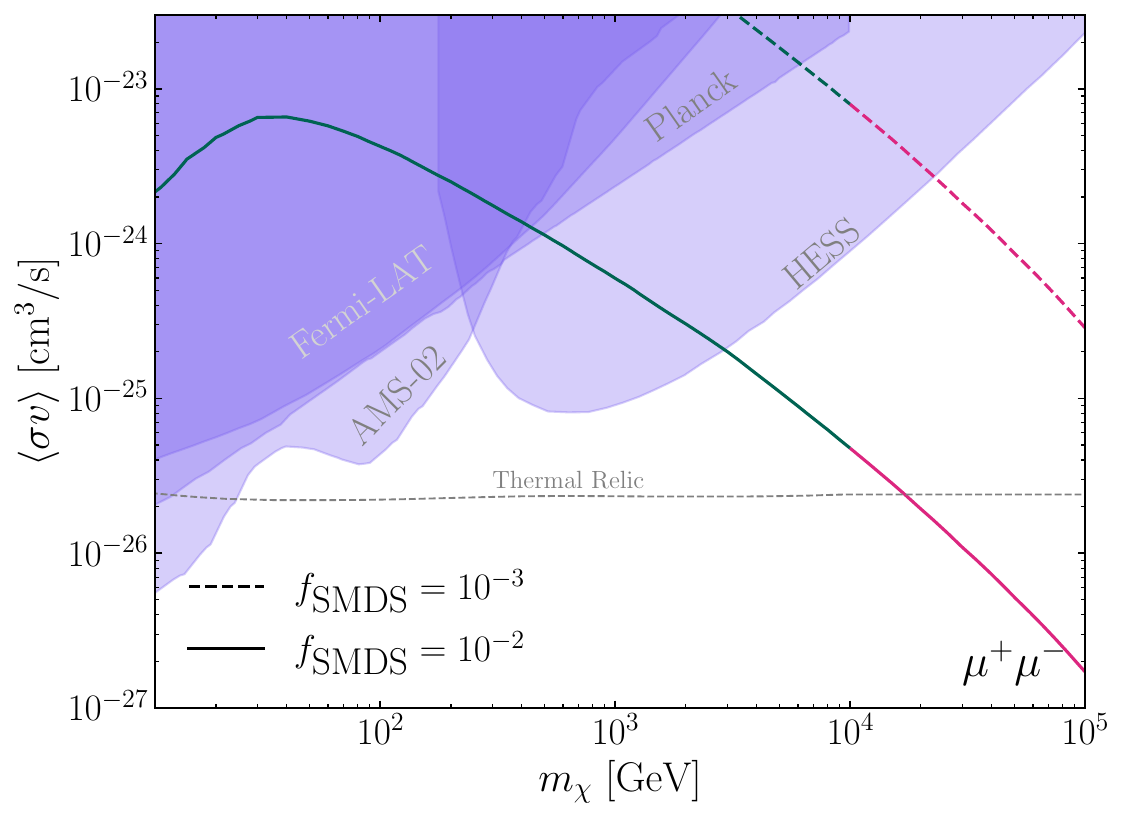}
    \caption{Same as Fig.~\ref{fig:DS_sens}, but for DM annihilation to $W^+W^-$ [Top Left], $b\bar{b}$ [Top Right] and $\mu^+\mu^-$ [Bottom]. Existing bounds from indirect DM detection searches by HAWC~\citep{HAWC:2023owv}, Fermi-LAT~\citep{Fermi-LAT:2015att}, AMS~\citep{AMS:2014xys, AMS:2014bun}, Planck~\citep{Planck:2018vyg} and HESS~\citep{HESS:2016mib} are shown as shaded regions.}
    \label{fig:sens}
\end{figure*} 

Besides electromagnetic diffuse emission, we can also compare DS emission with halo emission for DM annihilating to neutrinos.
 For comparison to the extragalactic neutrino flux from indirect DM detection, we are now interested in contributions from the entire halo population and compare to results of Ref.~\citep{Arguelles:2019ouk}. Noting that $f_\textrm{DS}$ varies by only $\mathcal{O}(1)$ factors for halo masses in the range of $10^6~M_\odot$ to $10^9~M_\odot$, we assume a constant value $f_\textrm{DS}\simeq 10^8$ for DSs that reach 1\% of the halo mass. Since only massive halos in the range $10^6-10^9~M_\odot$ can host supermassive DSss, this enhancement is suppressed.  Halos less than $10^6~M_\odot$ are far more abundant and we find that they contribute $\sim 10^3$ times more to the extragalactic neutrino flux than the heavier portion of the halo population. Thus $f_\textrm{DS} \sim 10^8$ enhances only portion of the total diffuse DS neutrino flux that is suppressed by a $\sim 10^{-3}$ due to subdominant contributions of large halos hosting DSs. From these approximate considerations, we expect supermassive DSs to enhance the neutrino flux compared to DM annihilation by a factor $\sim 10^5$. In Fig.~\ref{fig:extragalactic} we display our computational results, showing qualitative agreement with the estimates above and finding that the DS neutrino flux exceeds that of halos by around $\sim 10^6$.

\section{Different Dark Matter Annihilation Channels}\label{app:results}

In the main text we focused on results for DM annihilating to $\nu\bar{\nu}$. 
In Fig.~\ref{fig:sens}
we provide results for DM annihilation to $W^+W^-$, $b \overline{b}$ and $\mu^+\mu^-$ channels.  As in Fig.~\ref{fig:DS_sens} of the main text, we show  DM annihilation cross-sections for which neutrino emission from DSs in a fraction $f_\textrm{SMDS}$ of halos exceeds the atmospheric background at energies $E_\nu \gtrsim 200$~MeV. The total energy radiated in neutrinos is primarily determined by DS luminosity, which only weakly depends on DM annihilation channel. The difference in these results stems from the different neutrino spectra, with softer spectra such as that of $b\bar{b}$ channel being more challenging to distinguish and identify than the peaked spectra such as from $\nu\bar{\nu}$ channel.

 
\bibliography{refs}
\bibliographystyle{aasjournalv7}

\end{document}